# Ordered H$_2$O Structures on a Weakly Interacting Surface – A Helium Diffraction Study of H$_2$O/Au(111).

Authors: Gefen Corem[+‡], Pepijn R. Kole[†^‡], Jianding Zhu[^], Tatyana Kravchuk[+], J. R. Manson[◇] and Gil Alexandrowicz[+*].

AUTHOR ADDRESS

[+]Schulich Faculty of Chemistry, Technion - Israel Institute of Technology, Haifa 32000, Israel.

[^]The Cavendish Laboratory, J. J. Thomson Avenue, Cambridge, CB3 0HE, United Kingdom.

[◇]Department of Physics and Astronomy, Clemson University, Clemson, South Carolina 29634 USA.

ABSTRACT: In this manuscript we report helium atom scattering (HAS) measurements of the structure of the first $H_2O$ layer on Au(111). The interaction between $H_2O$ and Au(111) is believed to be particularly weak and conflicting evidence from several indirect studies has suggested that water either grows as 3D ice crystals or as an amorphous wetting layer. In contrast, our measurements show that between 110K and 130K, $H_2O$ grows as highly commensurate well-ordered islands which only partially wet the gold surface. The islands produce a clear diffraction pattern and are characterized by a well defined height of ~ 5Å with respect to the surface gold atoms. These findings provide support for a unique "double bi-layer" model which has recently been suggested for this surface.

**Introduction:**

In recent decades, numerous experimental and theoretical studies have been devoted to understanding the interaction of $H_2O$ with metallic surfaces[1-3]. This ongoing activity can be related to numerous technological applications, and the significant experimental and theoretical challenges encountered when studying these delicate and complex surface systems. In particular, the sensitivity of $H_2O$ molecules to electrons and radiation, the strong dependency of the observed structures on the exact preparation conditions and the difficulties encountered in theoretical studies have lead to conflicting results and controversy[1, 2, 4, 5].

An experimental method which played a leading role in determining the structure of water layers on metal surfaces is Low Energy Electron Diffraction (LEED)[1, 2]. While this technique has and is still supplying valuable information, it has a few limitations, in particular the sensitivity of water surfaces to electron induced damage and the lack of sensitivity to the positions of the protons within the layer, i.e., the molecular orientation of water. Significant progress has been made in recent years by combining state-of-the-art scanning tunneling microscopy (STM) and theoretical calculations (e.g., Refs.[ [3, 6, 7]]), a powerful combination which enables studying surfaces in real space and time. Real space techniques provide a major advantage when dealing with heterogeneous systems where the spatial averaging of diffraction techniques may lead to erroneous interpretation. On the other hand, STM studies of water surfaces are typically limited to ultra-low temperatures where the dynamics are extremely slow. Furthermore, stringent limitations on the scanning parameters are required to avoid significant perturbation effects, restricting these studies in terms of layer thickness, spatial resolution and the ability to relate apparent height to the real height [6, 8].

In this manuscript, we report helium atom scattering (HAS) measurements from an $H_2O$ overlayer adsorbed on an Au(111) surface. While HAS measures in reciprocal space which can complicate the structure analysis, especially on heterogeneous surfaces, it is an extremely non-perturbative technique, it is potentially very sensitive to light atoms, i.e., the proton order in a water layer (this topic will be discussed in more detail below) and it is capable of measuring the average structure on an atomic scale even in the presence of significant thermal motion. These properties make HAS particularly suitable for measuring $H_2O$ surface structures at elevated temperatures as has been demonstrated in the past, e.g., Refs. [ [9, 10, 11, 12]].

The Au(111) surface is believed to be characterized by a particularly weak $H_2O$-metal interaction, weaker than that of all the other close packed noble and transition metals[13, 14]. STM measurements at low temperatures and coverages have shown that $H_2O$ molecules predominantly populate the elbows of the herringbone reconstruction[8]. Isothermal desorption experiments suggested that water does not wet the Au(111) surface and that 3D growth of $H_2O$ takes place[15]. However, a couple of STM studies have since indicated the formation of a first layer which serves as a precursor to the growth of the subsequent layers[16, 17]. Furthermore, recent EELS measurements performed at 130K were also interpreted as indicating spread out ice layers[18] rather than 3D growth.

The current knowledge about the atomic-scale structure of the first layer of $H_2O$ on Au(111) is rather limited. Ikemiya and Gewirth performed STM measurements at 100K and concluded that the first layer of $H_2O$ on Au(111) is amorphous[16]. In contrast, a LEED pattern was reported by Pirug and Bonzel after dosing $H_2O$ at 120K[19].

It should be noted that patterns, expected from the classical hexagonal bi-layer model, were measured on several close-packed metal surfaces[1], however, some of these findings have later been related to the perturbation effects of the experimental techniques used[2, 3]. One relatively recent example is Pt(111) where previous work has identified structures[20], however, more recent STM experiments[21, 22] observe large unit cells and complex structures, consistent with HAS diffraction measurements[9]. In fact, during the last two decades experiments have detected a wealth of complex 1D, 2D and 3D water structures on metals as opposed to the expected classic bi-layer structures, raising doubts whether the hexagonal bi-layer model for water which was so widely embraced in the past is applicable to any $H_2O$ / pure-metal surface[4, 23].

In a recent combined STM, Infrared Adsorption Spectroscopy (IRAS) and Density Functional Theory (DFT) study, Stacchiola *et al*. studied the growth of the first $H_2O$ layer on Au(111) at 20K and at 100K[17]. Restrictions on the spatial resolution excluded direct measurements of the atomic scale structure or its symmetry, however, by combining STM apparent height measurements, DFT calculations and IRAS measurements the authors concluded that the first layer of $H_2O$ on Au(111) is a unique double-bilayer, i.e., two traditional proton ordered water bi-layers, one with a H-up and one with H-down stacked on top of each other. A schematic of the structure suggested by Stacchiola *et al.* is shown in the supplementary online information section. At elevated temperatures of ~ 100K the STM measurements were too blurry to analyze, however, measurements at 20K of films grown at 120K provided some insight, showing rather large islands (tens of nanometers large) that seem to nucleate at step edges.

**Experimental methods:**

The measurements presented below were performed using the Cambridge helium spin echo (HeSE) spectrometer[24] and a new HeSE instrument which is being constructed at the Technion. A description of the scattering parameters of the Technion setup which are relevant to this study can be found in the supplementary online information section. Both of these instruments were designed to perform ultra high energy resolution measurements of surface dynamics[25], but were used in this study as fixed scattering geometry helium diffractometers with a $45\text{Å}$ total scattering angle (using $^3$He at Cambridge and $^4$He at the Technion). Two Au(111) samples (Surface Preparation Lab, The Netherlands) were used in this study, both of which were aligned to within 0.1 degree and polished to a 0.03 micrometer roughness. The crystals were cleaned in-situ by repeated cycles of $Ar^+$ sputtering (0.8 kV) and annealing at 800 K until no further improvements in the helium specular reflectivity were observed. Reflectivity values > 15% were used as an indication for a clean and well ordered surface. Double distilled water was thoroughly degassed using freeze-pump-thaw cycles and was deposited on the surface using two different dosing methods: (1) a capillary array doser was located 15 cm from the sample[i] and (2) background gas was deposited from a leak valve. We have also varied the deposition flux between ~0.4L/min to 1 L/min (where L denotes Langmuir = $10^{-6}$ torr•sec), however, the diffraction data presented below did not seem to depend on any of these different deposition methods. The sample temperature was controlled using a liquid nitrogen cold finger and a filament located behind the sample holder. To assure that the water layer did not suffer any electron damage due to thermionic emission, we repeated the experiments while floating the sample at a negative potential (-50V), but did not observe any changes in the experimental results.

**Results and Discussion:**

Figure 1 shows the evolution of HAS angular scans along the <110> crystal azimuth as a function of the $H_2O$ coverage. The measurements were performed by monitoring the scattered helium intensity while rotating the surface normal within the scattering plane, and converting the scattering angle to the momentum transfer parallel to the surface, denoted as $\Delta K$. The sample temperature was 110K during both the overlayer growth and the HAS measurements. For the clean Au(111) we observe a sharp specular peak surrounded by the satellite peaks of the herringbone reconstruction, as was seen in previous HAS experiments[26], other diffraction features which were observed on the clean surface were weak first order peaks positioned along the <112> crystal azimuth (data not shown), which are also in agreement with the herringbone reconstruction pattern[26]. While exposing the surface to water, the intensity of the specular peak quickly decreases (the legend indicates the specular signal attenuation for each scan presented in the figure), and concurrently first and second order diffraction peaks emerge with a spacing of

---

i In order to calibrate the exposure, we performed experiments with both an indirect dose and the capillary array and determined the ratio of the pressure readings which give rise to a similar reduction of specular intensity.

ΔK=1.49±0.03 Å$^{-1}$ $^{ii}$. The intensity of these diffraction peaks initially increases, however, longer exposures lead to a reduction in intensity and eventually to the disappearance of the diffraction peaks (dose>10L), resulting in a broad featureless distribution as would be expected for an amorphous surface[10]. Figure 2 shows the evolution of the scattered signal from both the specular scattering condition and the first order peak along the <110> crystal azimuth as a function of $H_2O$ exposure. Two features which can be seen in the figure, are (1) a gradual increase of the diffraction signal with coverage, consistent with an island growth mode and (2) the existence of a maximum in the diffraction intensity, which occurs when the specular signal reduces by a factor of approximately 20 with respect to the intensity of the clean surface, the actual exposures needed to obtain the maximal diffraction signal intensity were sample dependent.$^{iii}$. A further point to note is that with increasing exposure beyond that at which the maximum in the diffraction intensity is reached, the intensities of the first- and second-order diffraction peaks decrease together at approximately the same rate. This rate is significantly smaller than the concomitant decrease in the specular intensity. That the attenuation rates of the first- and second-order diffraction peaks are similar is expected according to a Debye-Waller model for disorder[28] since the total momentum transfer is dominated by the perpendicular component which does not vary significantly between the two. In contrast, the much stronger attenuation of the specular peak suggests that its intensity is related to contributions from both the water-covered regions of the surface as well as bare Au(111) regions which coexist on the surface, an issue which we discuss in more detail below.

---

$^{ii}$ The two main contributions to the experimental uncertainty are the width of the energy distribution of the incident helium beam (0.2meV HWHM), and sample misalignment. A brief description of the method for measuring the energy distribution is given in the supplementary online information.

$^{iii}$ The exact dose which led to a specular attenuation of x20 and a maximum in the diffraction signal changed between 0.7L and 3L for the two different Au(111) crystals used in this study. The crystal which required larger exposures to achieve the x20 specular attenuation had a slightly lower specular reflectivity value of 15% (versus 20% for the second crystal), which could indicate a larger step density. Since the decoration of steps by adsorbates leads to a reduced specular attenuation with respect to covering flat terraces, larger exposures are needed to reduce the specular attenuation in comparison with a flat surface. Furthermore, step densities and their relation to co-adsorption phenomena have been shown to strongly influence $H_2O$ structure in a recent STM study on Pt(111)[27].        Standop, S.; Morgenstern, M.; Michely, T.; Busse, C., H2O on Pt(111): Structure and Stability of the First Wetting Layer. *J. Phys.: Condens. Matter* **2012**, 24, 124103.

Figure 3 shows a 2D angular scan obtained by repeating angular scans of the type shown in figure 1 for different crystal azimuths (i.e., rotating the sample about the surface normal between each angular scan) and plotting the logarithm of the measured scattered signal intensity. The temperature of the sample during the layer growth and measurement was 110K, the $H_2O$ exposure was stopped when the specular signal reduced by a factor of 20 corresponding to the optimal conditions of the diffraction peak mentioned above. In addition to the sharp specular peak surrounded by the herringbone satellite peaks (seen more clearly in the 1D measurements plotted in Figure 1), the measurement shows a well ordered diffraction pattern of the $H_2O$ overlayer.

The positions of the relatively sharp diffraction spots seen in our measurements are very close to those expected from a structure on the Au(111) surface (our measured peaks correspond to a structure which is compressed by less than 3% with respect to the bulk terminated surface), hence, given our experimental uncertainty the structure we measure can be considered as commensurate with the gold surface. Thus, as opposed to what one might expect, the weak $H_2O$-Au interactions do not prevent the growth of an ordered compact $H_2O$ structure which maintains a high degree of order and commensurability at relatively high temperatures. In fact, a further increase in the intensity of the first order diffraction peak intensity was observed when growing the layer at 130K (see figure S2 in the supplementary online information section). In contrast, the diffraction peak completely vanishes at 140K, a temperature which is close to the crystallization transition in bulk ice, and to a transition observed on Cu(111) where relatively flat $H_2O$ islands turn into corrugated pyramid like structures[6].

The fact that the herringbone reconstruction which characterizes the clean surface can be seen alongside the diffraction pattern could be interpreted as an indication that water does not lift the Au(111) reconstruction, however, the persistence of the pattern may also be simply related to the fact that the surface is only partially covered with water leaving exposed regions of reconstructed Au(111) surface. As explained below, the measurements presented in Figure 4 lead us to believe that the surface is in fact only partially covered under the conditions of our experiment.

Figure 4a shows the intensity of the helium specular reflection as a function of the perpendicular component of the wave vector which was scanned by heating the nozzle from 13.6K (2.9 meV) to 106K (23 meV) while maintaining specular conditions, i.e., without exchanging momentum within the surface plane. This type of measurement can be used to detect and study surface structures which exhibit order in the direction perpendicular to the surface plane, or alternatively to study the helium-surface interaction with high precision [29]. An overall trend seen for both the clean and $H_2O$ covered surface is an initial increase in intensity followed by a slow reduction at higher wave vector values, reflecting the source intensity dependencies and the Debye-Waller factor. Additional features are the oscillating patterns seen in both curves, a faint oscillation with a period of    for the clean Au(111) surface and a pronounced oscillation with a period of  after deposition of $H_2O$.

For specular conditions, an oscillation pattern is expected when the surface is characterized by flat regions with a well-defined height difference. In this case constructive interference is expected when the height between the two surfaces, , satisfies the relation . Therefore, the oscillations we observe on the clean surface correspond to a value of approximately , which is quite close to the expected height difference () between Au(111) terraces separated by a single-atom height step. The higher frequency seen in the red curve, indicates that the growth of $H_2O$ leads to interference between surfaces with a substantially larger height difference of ~. Figure 4b shows an alternative method of extracting the heights from the measurements. This plot shows the Fourier transform of the curves shown in Figure 4a, where the horizontal scale shows the angular frequency divided by 2 which should correspond to the height difference, . A faint peak at and a clear peak at can be seen for the clean and $H_2O$ covered surfaces respectively.

The data presented in Figure 4 indicates that under the same conditions where the optimal diffraction pattern is obtained we do not have complete wetting of the surface with a single water layer, as this would maintain the topography of the underlying gold surface and should not alter the oscillation period. This finding is consistent with the low temperature observations of partial dewetting on Cu(111)[6] given the fact that the $H_2O$-Au(111) interaction is assumed to be even weaker than that of $H_2O$-Cu(111)[14]. Hence, the oscillation pattern seen after water deposition, reflects interference effects between bare Au(111) regions and the surface of $H_2O$ islands which are characterized by a uniform height of approximately . This height value supports the proposition of a double bilayer $H_2O$ structure on Au(111)[17] and is approximately 20% lower than the DFT calculation performed on this system[30]. Our height value is also not too different (approximately 25% lower) with respect to the value deduced from STM measurements of the double bilayer structures observed on Cu(111)[6]. The fact that we do not observe larger-periodicity oscillations after exposing the surface to $H_2O$ indicates that if both single and double bi-layer islands coexist on the surface at our experimental conditions, the relative fraction of the single bi-layer islands is significantly smaller than that of the double bi-layer islands.

It is important to note that strictly speaking height estimations from oscillation curves should take into account the differences in the attractive part of the helium-surface interaction potential between the clean and the $H_2O$ covered part of the surface, as well as a shift of the repulsive part of the potential due to a different decay rate of the electron density into the vacuum. Unfortunately, we are not aware of any reliable potentials for $He/H_2O/Au(111)$ which would allow to perform such an analysis. Nevertheless, past experience shows that the errors expected from neglecting the details of the scattering potential[29] are not expected to be larger than the experimental uncertainty of our height values, which is dictated by the restricted number of oscillations we can measure and approximated as  . Finally, the lateral size of the ordered ice structures can also be estimated from the relatively narrow width of the 1st order diffraction peaks at the optimal growth conditions (FWHM<0.09 Å $^{-1}$), yielding an estimate of 70Å for the lateral extent of the well-ordered surface patches. We did not observe any significant change in this value after further (x3) $H_2O$ exposure even though further exposure significantly reduces the diffraction peak height[iv]. However, we did see a ~ 40% reduction in peak width, i.e., an increase in the size of the flat domains if the layer was grown at 130K (Figure S2 of the supplementary online information).

---

[iv] Such behavior would be expected if the additional $H_2O$ exposure covers some of the ordered water islands with an additional disordered layer which scatters diffusely and does not contribute to the diffraction pattern. In this case the residual diffraction signal reflects the diminishing number of islands which were still unaffected by the further growth and therefore maintain the same coherence length and corresponding peak width.

While the observations presented above indicate long range order with a particular symmetry and unit cell size, there are several questions which will need to be addressed in order to understand the exact real space structure of this system. One initial question is whether an ordered helium diffraction necessarily implies that the layer is proton ordered or not. On the one hand, HAS is not always guaranteed to detect proton order as has been shown by Glebov *et al*. for the surface of crystalline ice[10]. Glebov et al. showed that both a proton ordered model and an alternative model which includes a (relatively small) corrugation due to disordered protons, can be used to reproduce their experimental data. On the other hand, HAS is known to be very sensitive to light atoms and the structure of H-atoms on a clean surface has been studied in detail[31]. There is also growing evidence that HAS is sensitive to proton order in $H_2O$ surfaces. In particular, Gallagher *et al*. showed that the sensitivity of helium to proton order can explain the apparent discrepancy between the LEED and HAS results on $H_2O$/Ru(0001)[32]. Furthermore, evidence for the sensitivity of HAS to proton order has been shown in a recent study of a $H_2O$ layer on an oxygen precovered Ru(0001) surface[12], where the measured diffraction pattern was characterized by a glide line, demonstrating the sensitivity to the internal structure within the unit cell, i.e., the proton order. Thus, while HAS is definitely sensitive to hydrogen atoms, and can detect proton order on certain systems, the relation between a HAS diffraction pattern and proton-order needs to be assessed individually for any specific surface system. In particular, in systems of the type reported in this work, where there are no obvious missing diffraction spots indicating internal structure, further experimental support from other independent measurement techniques or detailed scattering calculations (when these can be reliably performed) are needed to relate the diffraction pattern and the proton order with confidence.

Regardless of whether the surface is proton ordered or not, the compact unit cell we observe is very different from the large unit cells seen on Pt(111) and Ni(111) [21, 33]. The compact commensurate structure we observe on Au(111), cannot be simply explained using geometrical considerations, as the gold surface does not provide better lattice matching than Pt(111) or Ni(111). Another interesting comparison is with Ru(0001), here one could naively expect that the close geometrical match and the strong interaction could lead to a well ordered bi-layer model. Nevertheless, previous measurements have shown that even though the oxygens adopt the pattern, the lack of a clear helium scattering diffraction pattern suggests a disordered surface[32]. It is possible that the mechanism which leads to the ordered structure and compact unit cell we observe on Au(111) is linked with the fact that the water islands seem to grow as two layer high structures, unlike what is seen on more reactive surfaces. The structural model introduced by Stacchiola *et al*. (figure S1 in the supplementary online information) provides a structure which is consistent with our measurements in terms of symmetry, commensurability and height, nevertheless, further experiments and/or elaborate scattering calculations are needed to validate this model, especially due to the considerable challenges encountered when performing DFT calculations of water surfaces in general and on Au(111) in particular[5, 34].

Common aspects can be found when comparing our results with STM studies on another weakly interacting system, $H_2O$/Cu(111), in particular, the partial dewetting mentioned earlier and the growth of double bilayer high $H_2O$ islands[6]. On the other hand, our measurements suggest long range order within the surface of the $H_2O$ islands, whereas the STM results, performed after annealing to 118K and cooling back to imaging temperatures, indicate an amorphous structure for the surface of the islands. Obviously these differences could be attributed to the variations between the copper and gold surfaces, however, they could also be a result of the different experimental probes used, in particular the different sensitivity to surface motion and perturbation effects for STM and HAS. Finally, ordered structures of $H_2O$/Cu(111) were seen with the STM when the surface was annealed to higher temperatures (140K)[6]. These pyramidal 4 layer-high structures developed at the temperatures of the amorphous to crystalline transition, and due to their heterogeneous appearance and highly corrugated surfaces would not be expected to provide a sharp helium diffraction of the type we measured. Hence, if these types of islands develop on Au(111) at similar temperatures, they might explain why we observed a disappearance of the diffraction pattern after annealing to 140K.

**Conclusions:**

In summary, using helium atom scattering we observe the formation of a water layer on Au(111). This layer is characterized by highly commensurate long-range ordered structures with a symmetry which grow as islands with a height of approximately . These findings provide support for the "double-bilayer" structure suggested recently for this system[17]. The ordered structures are observed at relatively high temperatures (up to 130K) and vanish only when the crystal is annealed at temperatures approaching the amorphous-to-crystalline transition of bulk ice. The high degree of order is surprising given other independent observations, in particular the notoriously weak Au-$H_2O$ interaction, the amorphous nature of $H_2O$ islands imaged with high resolution on Cu(111) and the amorphous structure observed on the much more reactive Ru(0001) surface[32]. One possible explanation is that the herringbone reconstruction of the Au(111) plays a special role in establishing order within the $H_2O$ layers, other explanations might be due to the different experimental probes used. Further comparative studies are needed to answer these questions.

FIGURES

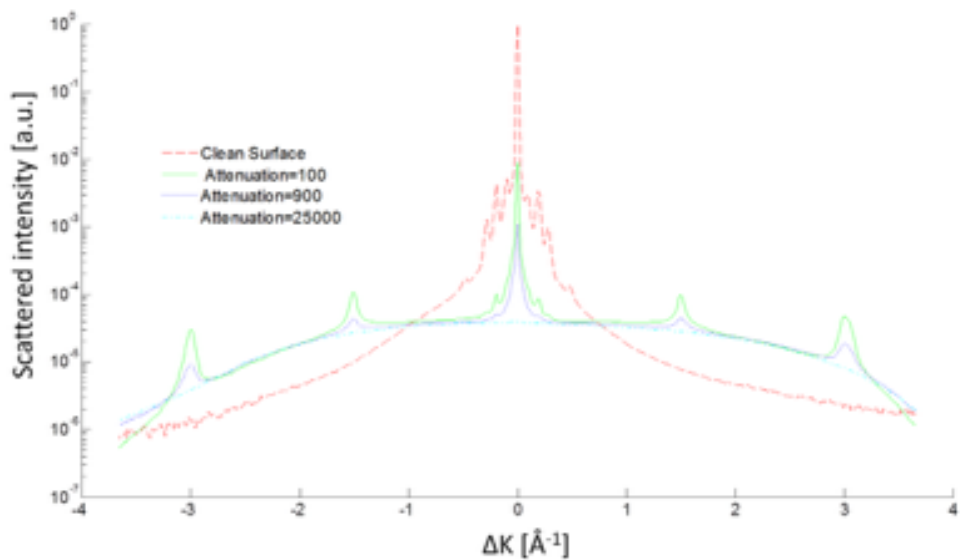

Figure 1. Angular scans measured at 110K along the <110> crystal azimuth using an 8 meV $^3$He beam. The measurements show the evolution of the angular distribution with increasing doses of $H_2O$, from a clean Au(111) surface (red dashed line), through an ordered $H_2O$ over-layer (green full line, and blue dotted line) and up to what seems like an amorphous film of $H_2O$ (dash-dot cyan line), obtained after substantial $H_2O$ exposure (>10L) .

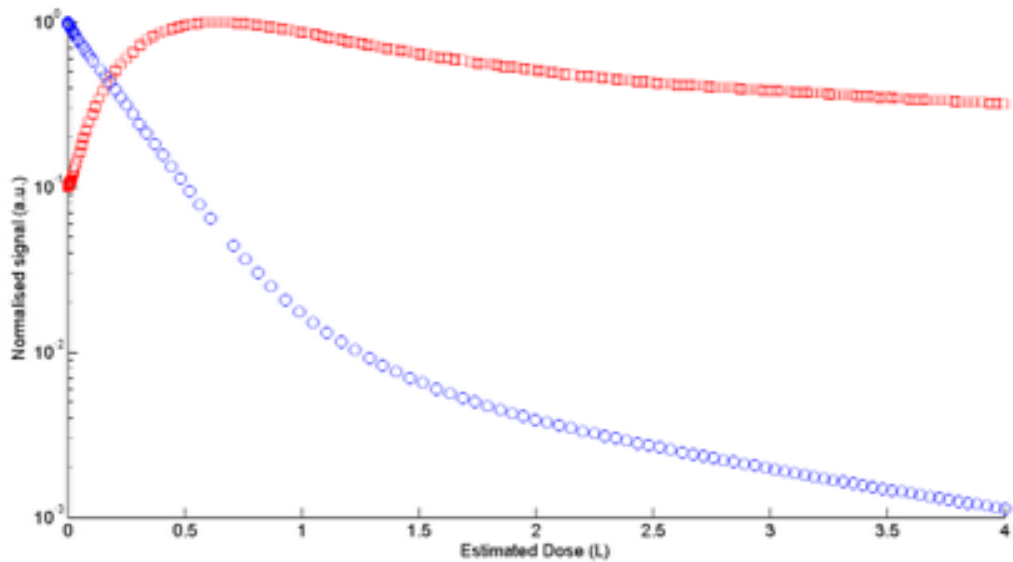

Figure 2. Evolution of the scattered helium intensity as a function of $H_2O$ exposure measured at 110K. The normalized specular intensity is plotted using blue circular markers, and decays continuously as a function of exposure. The normalized intensity at the first order diffraction pattern along the <110> crystal azimuth is plotted using red square markers. The diffraction signal increases gradually with exposure and peaks at an exposure which attenuates the specular intensity by ~ x20. Further exposure reduces the diffraction signal intensity, indicating the growth of a multilayer amorphous solid water layer.

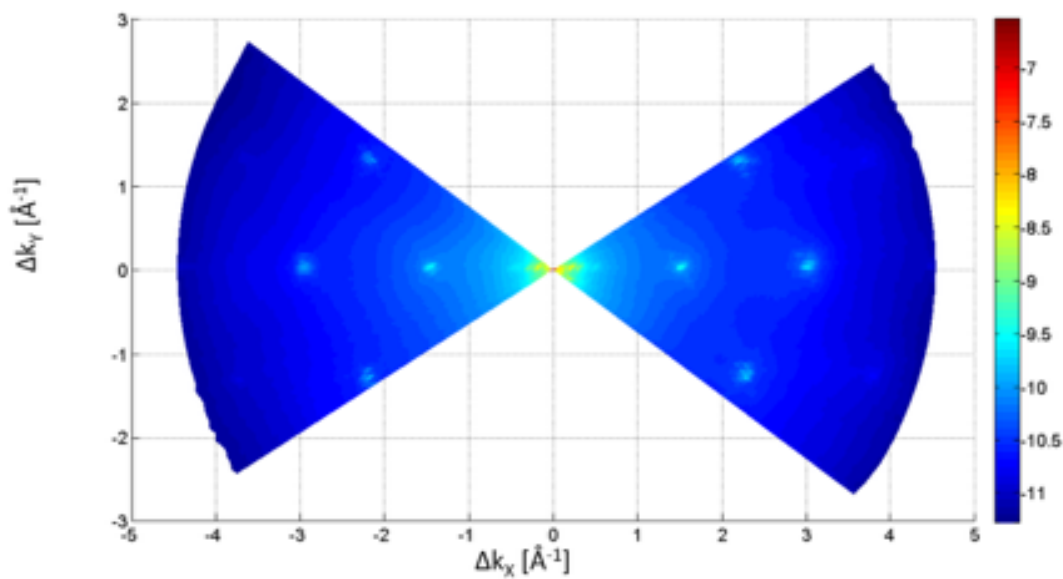

Figure 3. 2D color plot of the logarithm of the scattered helium intensity as a function of parallel momentum transfer performed using a 12 meV $^3$He beam. The $H_2O$ layer was grown at 110K under the optimal exposure conditions described in the text. The image was obtained by combining 71 in-plane 1D angular scans (of the type shown in Figure 1) performed for different crystal azimuth angles. The <110> azimuth is oriented along the horizontal axis of the plot. Eight strong and sharp diffraction features can be seen in the high symmetry directions, whereas additional weaker peaks can be resolved for larger momentum transfer values in between the high symmetry directions. In addition to the characteristic herringbone peaks seen close to the origin ($\Delta K=0$), there seems to be some very faint satellite peaks surrounding the main diffraction peaks along the diagonal azimuth, while the limited resolution and signal to noise ratio of the measurement do not allow a detailed analysis of these peaks, their position is also in agreement with the expected herringbone reconstruction pattern[26].

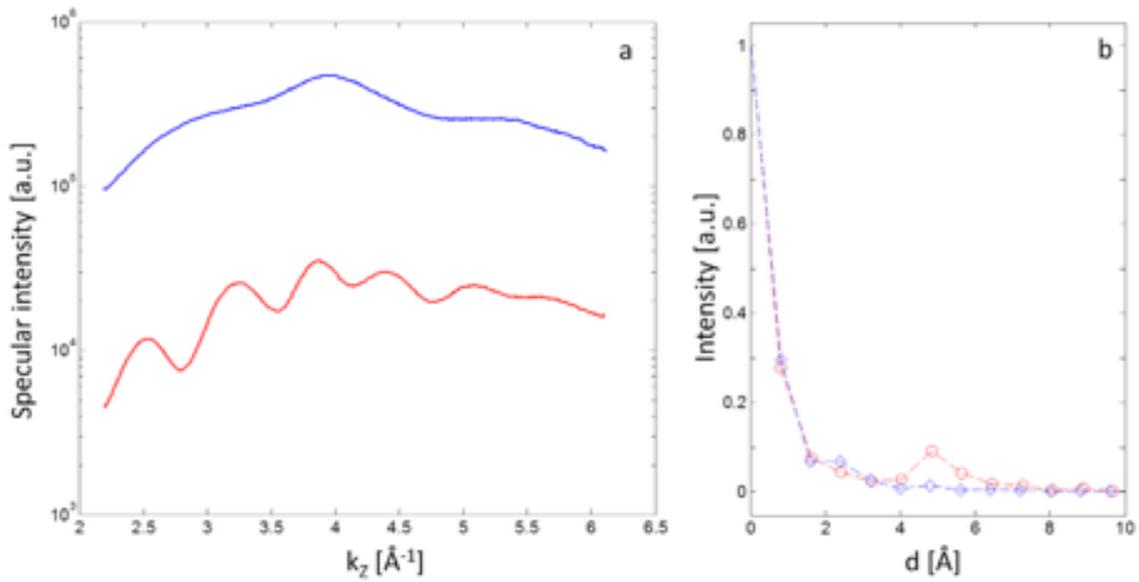

Figure 4. (a) Specular signal intensity as a function of the perpendicular component of the helium beam wave vector. Measurements were performed by changing the energy of a $^4$He beam between 2.9 and 23 meV. The surface temperature for both the clean Au(111) (blue) and the $H_2O$/Au(111) (red) measurements was 110K, the $H_2O$ layer was produced using the optimal diffraction conditions mentioned in the text above. (b) Normalized Fourier transforms of the curves shown in the left panel (blue diamonds and red circles correspond to clean and $H_2O$ covered surfaces). The horizontal scale plots the angular frequency divided by two, which corresponds to the height difference, , between the interfering surfaces. The resolution within which $d$ can be determined is governed by the total range of $k_z$ values measured in the experiment.

AUTHOR INFORMATION

**Corresponding Author**

* Schulich Faculty of Chemistry, Technion - Israel Institute of Technology, Technion City, Haifa 32000, Israel. Email: ga232@tx.technion.ac.il

**Present Addresses**

† Shell Upstream International, Schepersmaat 2, 9405 TA Assen, The Netherlands

**Author Contributions**

The manuscript was written through contributions of all authors. All authors have given approval to the final version of the manuscript. ‡These authors contributed equally.

**Funding Sources**

This work was funded by the European Research Council under the European Union's seventh framework programme (FP/2007-2013)/ ERC Grant 307267, the United States-Israel Binational Science Foundation - Grant 2010095 and the German-Israeli Foundation for Scientific Research and Development.

ACKNOWLEDGMENT

The authors would like to thank Dr. J. Ellis and Dr. W. Allison from the Cavendish Laboratory for generously providing access to their spectrometer, Prof. H. Ibach for helpful discussions and Dr. Ping Liu and Dr. Dario Stacchiola from the Brookhaven National Laboratory who kindly provided us the details of their DFT results. One of us (JRM) would like to acknowledge the generous hospitality of the Chemistry Department of the Technion.

Supporting Information Available: A description of the double bi-layer model, experimental details and a comparison of the diffraction peak intensity at different growth temperatures is available free of charge via the Internet at http://pubs.acs.org.

# Supplementary online information:

Title: Ordered H$_2$O structures on a weakly interacting surface – a helium diffraction study of H$_2$O/Au(111).

Authors: Gefen Corem[+‡], Pepijn R. Kole[†^‡], Jianding Zhu[^], Tatyana Kravchuk[+], J. R. Manson[<>] and Gil Alexandrowicz[+*].

AUTHOR ADDRESS

[+]Schulich Faculty of Chemistry, Technion - Israel Institute of Technology, Haifa 32000, Israel.

[^]The Cavendish Laboratory, J. J. Thomson Avenue, Cambridge, CB3 0HE, United Kingdom.

[<>]Department of Physics and Astronomy, Clemson University, Clemson, South Carolina 29634 USA.

1.1 The "double bi-layer" model

As explained in the manuscript a unique "double bi-layer" model was recently suggested for H$_2$O/Au(111) by Stacchiola and co authors[1], a model which is consistent with the experimental observations we presented in the manuscript. The "double bi-layer" model was suggested based on a combination of IR, STM and DFT work, and is characterized by a (H-up) bi-layer on top of the gold surface covered by a second (H-down) bi-layer above it. The unit cell of the surface layer is commensurate with the gold substrate and has a $(\sqrt{3} \times \sqrt{3})R30$ unit cell. The uppermost nuclei in this "double bi-layer" structure are positioned 6.2 Å above the gold surface atoms. In order to assist the readers in visualizing this model we have drawn a schematic of the Stacchiola *et al.* model in figure S1, based on the DFT nuclei coordinates kindly supplied to us by Dr. Ping Liu from the Brookhaven National Laboratory.

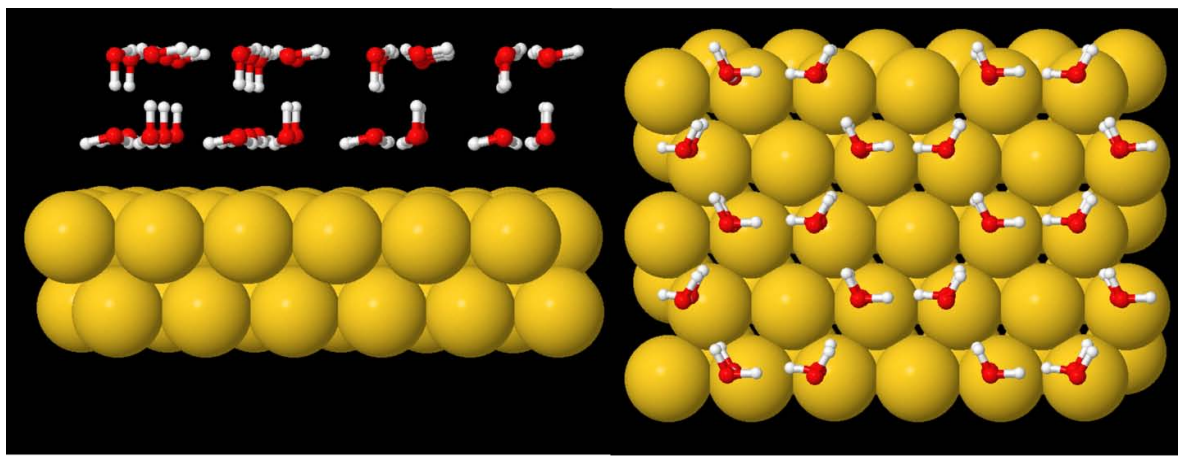

**Figure S1. Side and top views of the "double bi-layer" suggested by Stacchiola *et al.* and plotted using their DFT calculated nuclei coordinates. The image was created using Jmol: an open-source Java viewer for chemical structures in 3D.**

1.2 Details of the helium scattering apparatus:

As mentioned in the text, two helium spin echo (HeSE) instruments were used to measure the presented data, these instruments were designed to study surface dynamics but were used as helium diffractometers for the purpose of this study. The Cambridge HeSE spectrometer was used to measure the data presented in figures 1-3 of the manuscript, and the details of this instrument have been published previously[2]. Figures 4 and S2 were measured using a new HeSE which is currently being developed at the Technion. The main parameters which describe the instrument at the time the data presented in this manuscript was measured, are given in table 1.

**Table 1**

| Beam gas | $^4$He |
|---|---|
| Nozzle diameter | 40μm |
| Stagnation pressure | 1 bar |
| Incident energy | 2.9 meV – 23 meV |
| Beam diameter | 1.5 mm |
| Angular resolution at detector | 0.2° |

| Total Scattering angle | 45° |
|---|---|
| Sample manipulator and temperature control. | Home built 6 axis manipulator with sample cooling (liquid nitrogen/helium) and heating (radiative and electron bombardment) capabilities with a range of 60K to 1100K and a stability < ±0.2K . The sample was mounted on a homebuilt sample cartridge designed by Dr. Andrew Jardine. |
| Detector | Quadrupole (Extrel-MAX 120) |
| Scattering chamber base pressure | P<5x10$^{-10}$ mbar |

1.3 Calibration of the momentum transfer.

When analyzing the diffraction data shown in the manuscript, the precision of the calculated momentum transfer for the diffraction peaks impacts on the inferred structure and on the level of commensurability with the underlying substrate. We estimate the accuracy of our results as $\Delta K=\pm 0.03$ Å$^{-1}$. The contributing factors are the uncertainty in the scattering angles due to imperfect sample alignment and an uncertainty in the incident energy related to the spread of energies in the incident beam. In order to provide accurate values for the momentum transfer it is important to calibrate the incident beam energy and measure the fixed total scattering angle defined by the relative positions of the detector and the source. The calibration procedure of the Cambridge apparatus involved spin-rotation measurements (horizontal tilted projection measurements[3]) from a LiF sample, used as a monochromator with a well known diffraction pattern, details of the procedure have been described in the PhD Thesis of S. Dworski[4]. The Technion apparatus did not have operational spin rotation coils during the time of the measurements presented in this manuscript and the incident energy was estimated using the relation for the terminal velocity of a mono-atomic supersonic beam $E=(5/2)k_BT$, this estimation proved to be sufficiently accurate as the diffraction peak positions we measured using the two instruments differed by no more than 0.01Å$^{-1}$.

1.4 Effect of temperature on diffraction pattern.

Figure S2 compares angular scans of an $H_2O$ layer which were deposited at three different temperatures, the upper panel at 110K, the middle panel at 130K and the lower panel at 140K. All three measurements were performed at 110K. As can be seen, growing at 130K gives a stronger peak which is also sharper, indicating that at this temperature the flat ordered domains are larger. Heating the surface further to 140K leads to a complete loss of the diffraction pattern as discussed in the text.

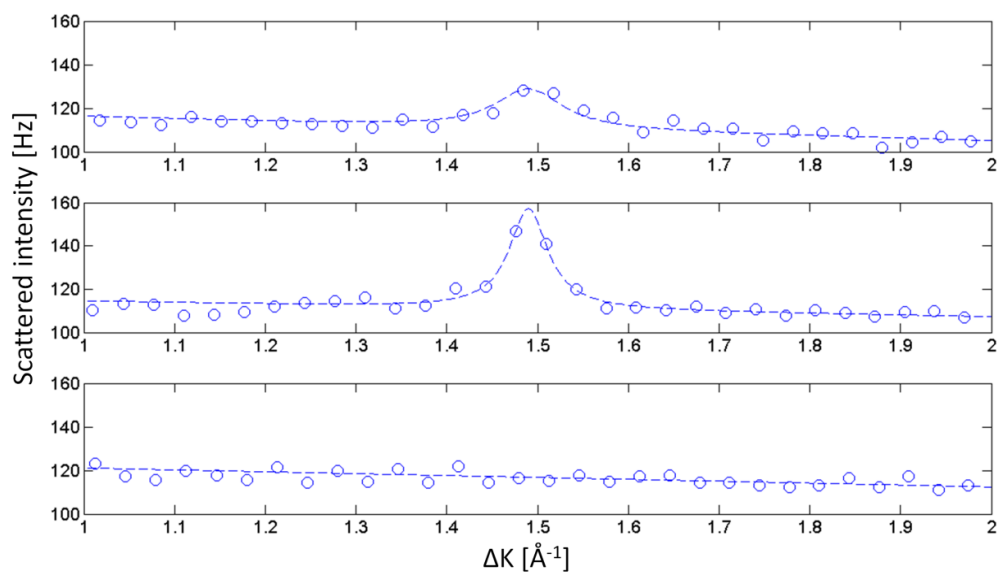

**Figure S2.** Comparison of angular scans measured along the <110> crystal azimuth of a $H_2O$ layer grown at 110K (top), 130K (middle) and 140K (bottom). For all three temperatures, the surface was exposed to $H_2O$ until the specular signal reduced by x20 and the angular measurements were performed at 110K. The dashed line is a fit to a Lorentzian peak centered at 1.48 Å$^{-1}$ superimposed with an additional linear term to fit the overall reduction in intensity between 1 and 2 Å$^{-1}$.